\documentclass[3p,times,twocolumn]{elsarticle}

\usepackage{ecrc}

\volume{00}

\firstpage{1}

\journalname{Nuclear Physics B Proceedings Supplement}

\runauth{A. Kappes}

\jid{nuphbp}

\jnltitlelogo{Nuclear Physics B Proceedings Supplement}

\usepackage{amssymb}

\usepackage{lineno}

\usepackage[figuresright]{rotating}

\usepackage[gen]{eurosym}

\begin{document}

\begin{frontmatter}



\dochead{}

\title{Exploring the Universe with Very High Energy Neutrinos}


\author[kappes]{A. Kappes}
\author[icecube]{for the IceCube Collaboration}

\address[kappes]{Erlangen Centre for Astroparticle Physics, Friedrich-Alexander-Universit\"at Erlangen-N\"urnberg, D-91058 Erlangen, Germany}
\fntext[icecube]{http://icecube.wisc.edu}

\begin{abstract}
With the discovery of a high-energy neutrino flux in the 0.1\,PeV to PeV range from beyond the Earth's atmosphere with the IceCube detector, neutrino astronomy has achieved a major breakthrough in the exploration of the high-energy universe. One of the main goals is the identification and investigation of the still mysterious sources of the cosmic rays which are observed at Earth with energies up to several $10^5$ PeV. In addition to being smoking-gun evidence for the presence of cosmic rays in a specific object, neutrinos escape even dense environments and can reach us from distant places in the universe, thereby providing us with a unique tool to explore cosmic accelerators. This article summarizes our knowledge about the observed astrophysical neutrino flux and current status of the search for individual cosmic neutrino sources. At the end, it gives an overview of plans for future neutrino telescope projects.
\end{abstract}


\end{frontmatter}


\section{Introduction}
\label{}
The discovery of cosmic rays by Victor Hess in 1912 marked the beginning of astroparticle physics which has enabled us to explore the Universe at the highest energies reaching up to several $10^{5}$\,PeV. But even more than 100 years later, the central question regarding the sources that can accelerate particles to energies far beyond what is achievable with man-made accelerators is mostly unanswered. First of all, cosmic rays cannot directly reveal their sources except maybe at the very highest energies as the charged particles are deflected in the galactic and intergalactic magnetic fields and hence do not point back to their origin. High-energy gamma-ray photons, which are produced in the interaction of cosmic rays with matter or photon fields, on the other hand, can also be generated via up-scattering of low-energy photons by accelerated electrons (inverse Compton scattering). Though the gamma-ray spectra of at least two supernova remnants seem to originate from pion decay and hence from acceleration of protons or heavier nuclei \cite{sc:339:807}, a clear picture whether supernova remnants are indeed the main sources of the Galactic cosmic rays is still missing. Furthermore, the origin of ultra-high energy cosmic rays (UHECRs) above $\gtrsim 3\times 10^{18}$\,eV remains a complete mystery. With their unique properties, neutrinos will help in solving these and other important astrophysical questions, as their observation from an object or region would unambiguously identify it as a source of high-energy protons or heavier nuclei. This has been one of the main drivers for the development and construction of neutrino telescopes over the past decades. At the same time, the fact that until recently the high-energy neutrino sky has been total terra incognita implies a high potential for unexpected discoveries. 

The low interaction cross section that renders neutrinos very valuable for the exploration of the high-energy universe, however, also makes them very hard to detect. In particular, over the past years it has become clear that huge detectors of at least km$^3$ size are necessary to eventually open this exciting new window to the universe (see e.g.\ \cite{apj:656:870,app:34:778}). \vfill

\begin{figure*}[t]
\includegraphics[height=5cm]{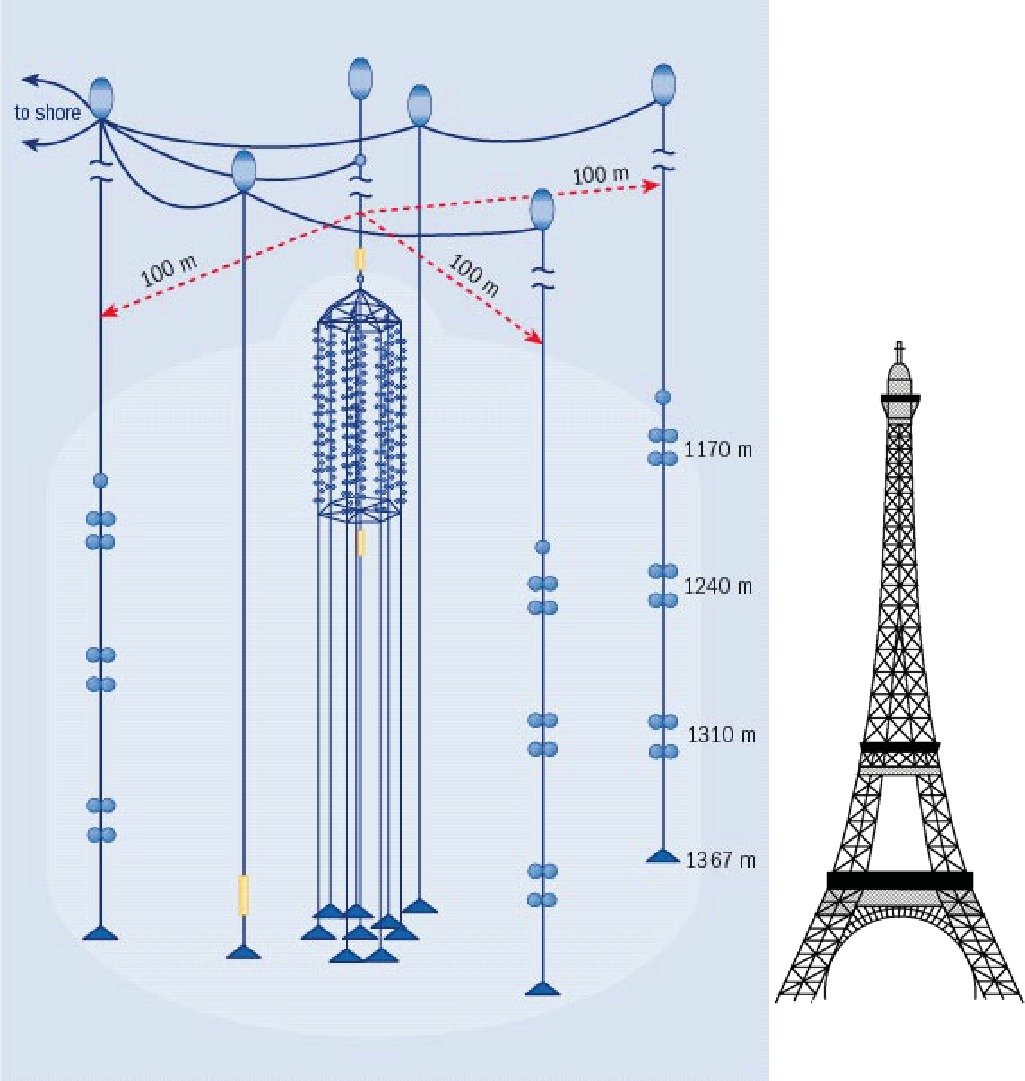}
\hfill
\includegraphics[height=5cm]{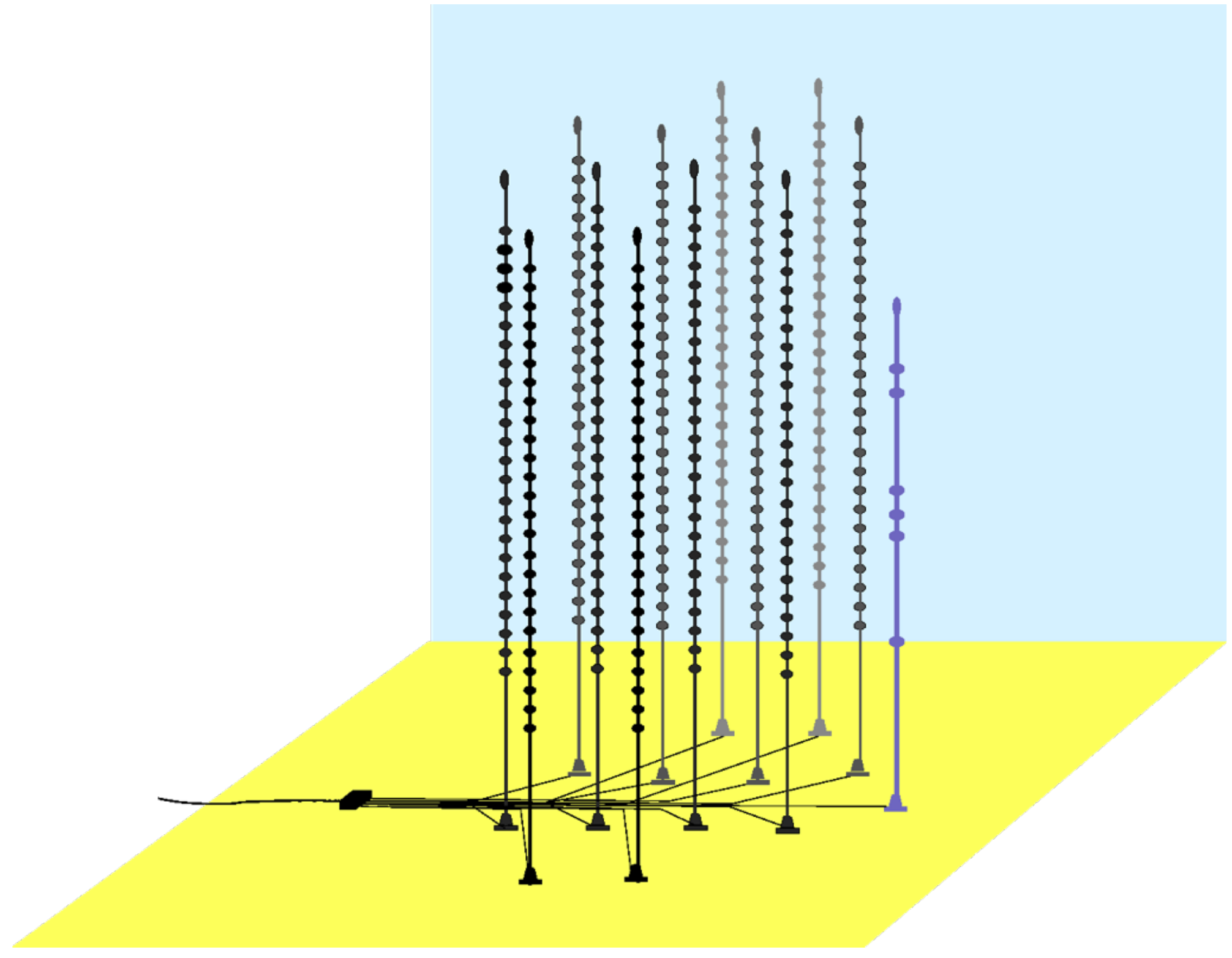}
\hfill
\includegraphics[height=5cm]{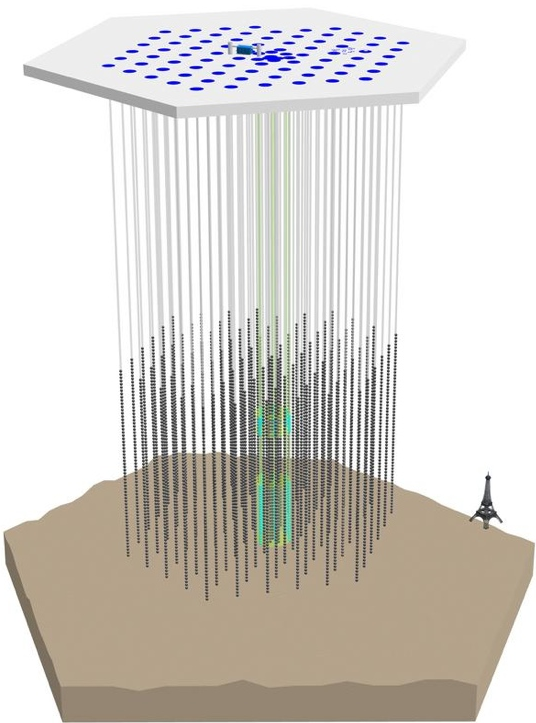}
\caption{Schematic views of the BAIKAL (left) \cite{app:7:263}, ANTARES (middle)\cite{nim:a656:11} and IceCube (right) \cite{nim:a700:188} neutrino detectors. The Eiffel Tower is shown for scale comparison.}
\label{fig:detectors}
\end{figure*}

\section{Neutrino telescopes}
The idea for neutrino telescopes was first published in 1960 by Markov \cite{proc:airc1960:578}. The detection principle is based on the registration of the Cherenkov light induced by charged particles generated in neutrino interactions in an optically transparent medium like ice or water. This light is recorded with a large number of photomultipliers arranged in a three-dimensional array. The direction and energy of the neutrino is reconstructed using the arrival time of the photons (measured with nanosecond precision), the measured light intensity and the position of the photomultipliers. 

The technical realization of such a telescope in deep ocean water was pioneered by the DUMAND Collaboration between 1973 and 1995 \cite{rmp:64:259} but terminated after a technical failure of the first deployed string. In the 1980's, the construction of detectors in Lake Baikal \cite{app:7:263} and in the ice at the South Pole were proposed. The NT200 in Lake Baikal was completed in 1998, instrumenting a volume of $10^{-4}\,\mathrm{km}^3$ with 192 optical modules on eight strings (Fig.~\ref{fig:detectors}, left). Three additional strings at larger distances were added in 2005--2007. The AMANDA detector at the South Pole \cite{na:410:441} took data from 1996 until 2009. In its final configuration it consisted of 667 optical modules instrumenting a volume of about $10^{-2}\,\mathrm{km}^3$. Installation of a neutrino telescope in the deep ocean was pursued by the ANTARES, NEMO and NESTOR Collaborations in the Mediterranean Sea. The ANTARES detector (Fig.~\ref{fig:detectors}, middle) was eventually built off the coast of southern France near Toulon \cite{nim:a656:11} with construction lasting from 2002 to 2008. It comprises 885 optical modules on 12 strings and instruments a volume of $10^{-2}\,\mathrm{km}^3$. Data taking started early in the construction phase and is ongoing. In 2005, construction of the IceCube detector (Fig.~\ref{fig:detectors}, right), the successor of AMANDA, started with the aim to build the first km$^3$-scale neutrino telescope \cite{nim:a601:294}. In its final configuration, reached in 2010, the detector consists of 5160 optical modules instrumenting one $\mathrm{km}^3$ of clear glacial ice at depths between 1450\,m and 2450\,m at the geographic South Pole. Physics data taking started in 2006 with 9 installed strings. In contrast to the other detectors, the IceCube Observatory comprises an air shower array at the surface called IceTop \cite{nim:a700:188}. Its main purpose is the investigation of cosmic rays in the energy range between $10^{15}$\,eV and $10^{18}$\,eV \cite{prd:88:042004}. This article concentrates on results from the IceCube and ANTARES detectors as these are currently the most sensitive detectors in the Southern and Northern hemisphere, respectively.

\begin{figure}[t]
\begin{center}
\includegraphics[width=5cm]{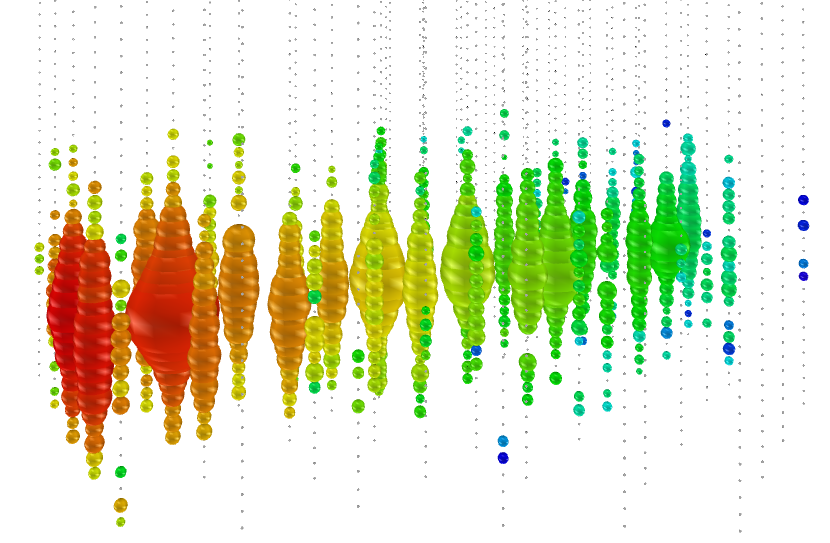}
\\[5mm]
\includegraphics[width=5cm]{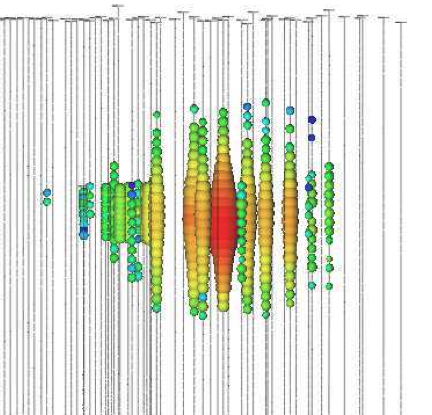}
\end{center}
\caption{Neutrino interaction signatures in a neutrino telescope (here, two events from IceCube data). The size of the colored spheres is proportional to the number of detected photons in an optical module and the color indicates the arrival time of the first photon on that module, with red being early and blue late times. Top: Track-like pattern from a muon originating from a charged current interaction of a muon neutrino. Bottom: cascade-like pattern from e.g.\ a neutral current neutrino interaction. }
\label{fig:signatures}
\end{figure}

Two basic event topologies in neutrino telescopes can be distinguished. Charged-current interactions of muon neutrinos produce long track-like patterns due to the resulting muon crossing the detector (muon channel; Fig.~\ref{fig:signatures}, top). On the other hand, neutral-current interactions of all neutrino flavors, or charged current interactions of electron neutrinos, induce a more spherical hit pattern originating from the cascade of particles produced at the interaction vertex (cascade channel; Fig.~\ref{fig:signatures}, bottom). In case a charged current muon neutrino or tau neutrino interaction with subsequent tau decay into a muon happens inside the detector these two topologies overlap thereby complicating the reconstruction. The direction of elongated muon tracks can be reconstructed with sub-degree precision at high energies, significantly better than that of cascades. At the relevant energies, the neutrino is approximately collinear with the muon and, hence, the muon channel is the prime channel for the search for point-like sources of cosmic neutrinos. On the other hand, cascades deposit all of their energy inside the detector and therefore allow for a much better reconstruction of their energy with a resolution of up to 10\% at high energies.

\section{Atmospheric muon and neutrino fluxes}
The main background for the detection of cosmic neutrinos originates from muons and neutrinos produced in interactions of cosmic rays with Earth's atmosphere. Atmospheric muons are commonly suppressed by looking through the Earth to the opposite hemisphere using the Earth as an absorber. However, for energies above $\sim$100\,TeV the Earth starts to become opaque also for neutrinos which renders the upward direction the only part of the sky from which neutrinos with EeV energies and above can be observed. Fortunately, at these energies the atmospheric muon background is negligible. 

\begin{figure}[t]
\includegraphics[width=7cm]{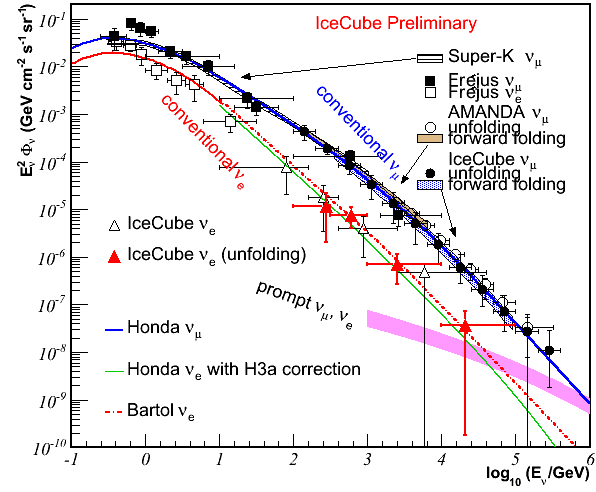}
\caption{Atmospheric neutrino fluxes as a function of energy. Shown are predictions for the conventional electron and muon neutrino fluxes together with the respective measurements. Predictions for the prompt neutrino fluxes are also shown. This plot is based on the plot published in \cite{prl:110:151105} with an update to the IceCube measurement of electron neutrinos displayed in red.}
\label{fig:atm_nu}
\end{figure}

Atmospheric neutrinos are a tough background for all searches for cosmic neutrinos. On the other hand, they are invaluable as a calibration source and also allow to investigate very interesting physics topics like for example neutrino oscillations \cite{pl:b714:224,prl:111:081801}. Figure~\ref{fig:atm_nu} shows the atmospheric muon and electron neutrino fluxes as measured by several experiments, together with theoretical predictions. The \emph{conventional} neutrino flux \cite{pr:d75:043006} originates from the decay of kaons and pions produced in cosmic-ray interactions in the atmosphere ($\pi^\pm,K^\pm \rightarrow \mu^\pm + \nu_\mu$) and the subsequent muon decay ($\mu^\pm \rightarrow e^\pm + \nu_\mu + \nu_e$). Electron neutrinos are less abundant than muon neutrinos as they are only generated in muon decays. The $\nu_e / \nu_\mu$ ratio decreases with increasing energy as an increasing number of muons reaches the detector before decaying. At energies above $\gtrsim 100$\,TeV, electron neutrinos from the decay of $K_L$ and $K_S$ constitute a significant fraction of the total electron neutrino flux \cite{arxiv:1409:4924}. Measurements of the conventional neutrino fluxes cover a broad energy range up to energies of several 100\,TeV and are well described by theoretical predictions. The \emph{prompt} neutrino flux stems from the decay of charmed particles generated in the very early phase of the air shower development. As these particles are very short-lived they decay without interaction and therefore yield a harder spectrum compared to the conventional neutrino flux, with about equal fluxes of muon and electron neutrinos. Currently, there exists no measurement of this prompt flux and its normalization has large theoretical uncertainties \cite{pr:d78:043005}. 

\section{The classical picture of neutrino astronomy}
Compared to atmospheric muons and neutrinos, cosmic neutrinos are expected to have a harder spectrum following a power-law with an index of about $-2$ as inferred from shock acceleration (see e.g.\ \cite{aaa:a34:553} and references therein) in objects like supernova remnants or jets of active galactic nuclei. Therefore, cosmic neutrinos are expected to dominate over the atmospheric background above a certain energy threshold. In order to lower this energy threshold into the TeV to PeV range, where in particular Galactic sources are anticipated to produce neutrinos, ''classical'' neutrino astronomy uses the Earth as shield against the overwhelming flux of atmospheric muons.

\subsection{Searches for point-like sources}
\begin{figure}[t]
\includegraphics[width=7.5cm]{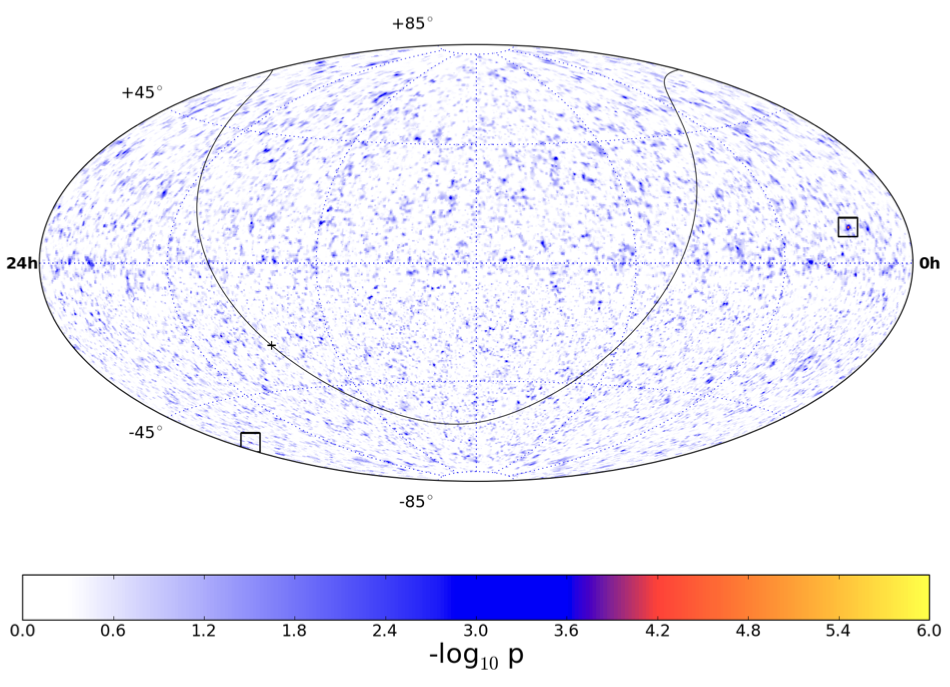}
\caption{IceCube pre-trial significance skymap in equatorial coordinates. The black line indicates the Galactic plane, and the black plus sign indicates the Galactic Center. The most significant fluctuation in each hemisphere is indicated with a square marker. Taken from \cite{arxiv:1406.6757}.}
\label{fig:skymap_icecube}
\end{figure}

\begin{figure}[t]
\includegraphics[width=7.5cm]{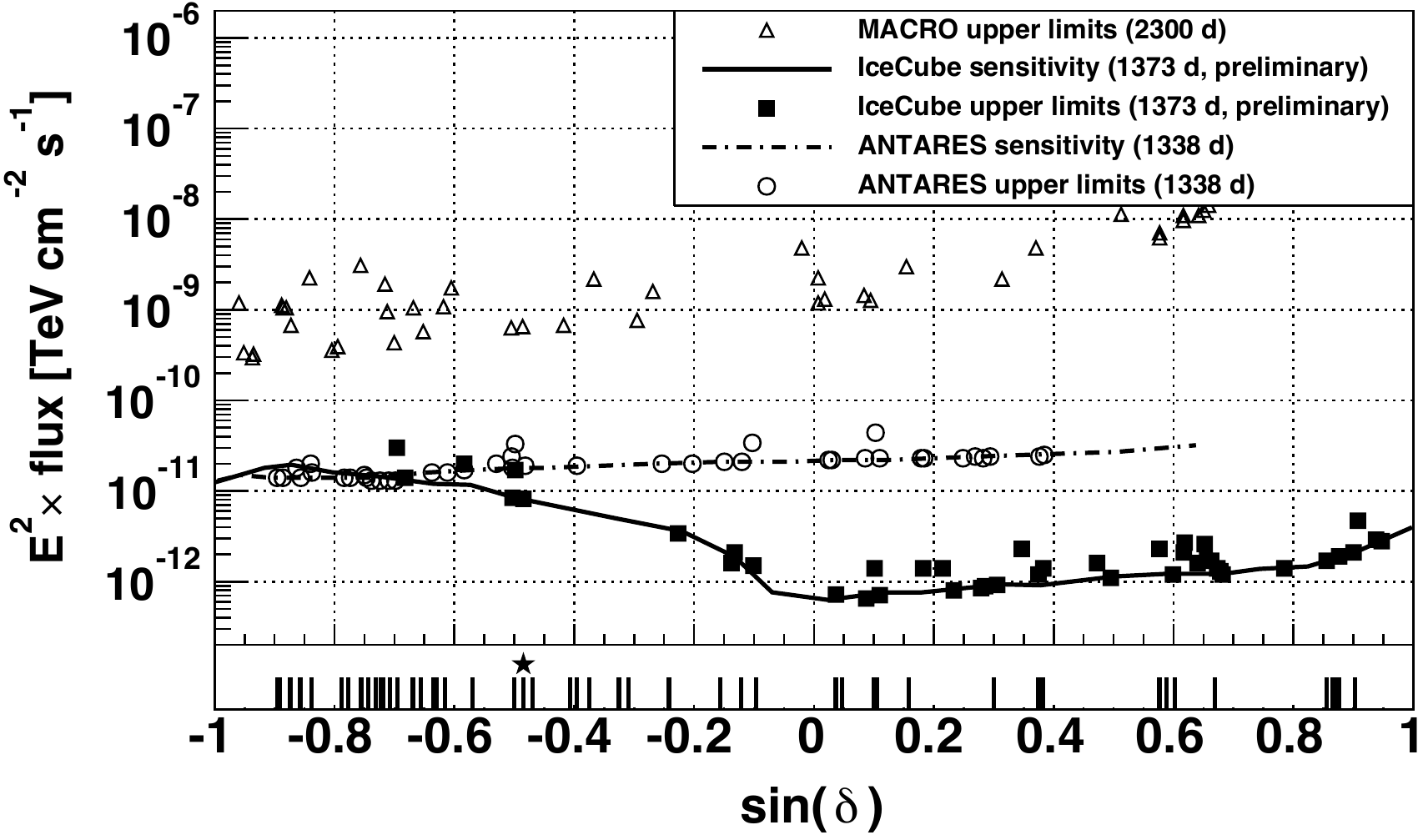}
\caption{Sensitivities (solid lines) and upper limits (symbols) of various experiments in myon neutrinos for point-like sources as a function of declination assuming an unbroken $E^{-2}$ spectrum. Data taken from \cite{arxiv:1406.6757,apj:760:53,apj:546:1038}.}
\label{fig:ps_limits}
\end{figure}

\begin{figure}[t]
\includegraphics[width=7.5cm]{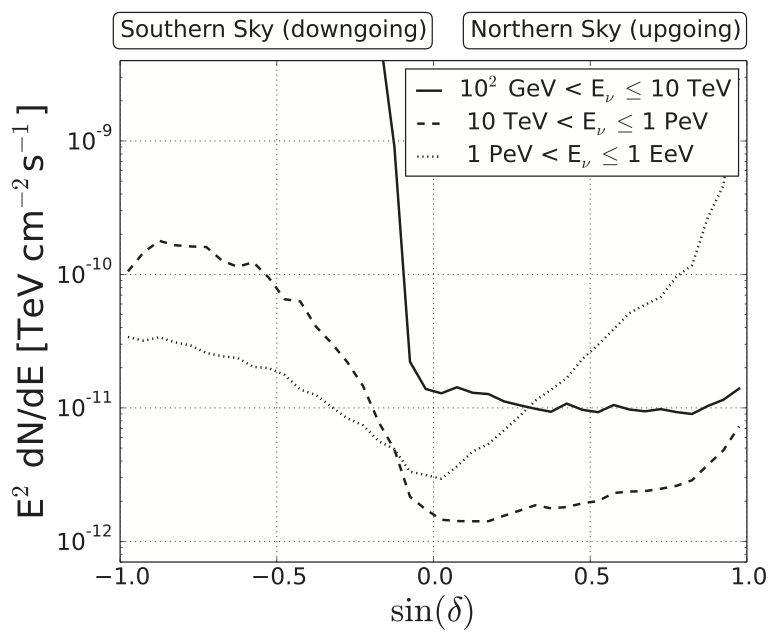}
\caption{Sensitivity of the IceCube detector in muon neutrinos to a point-like sources with an $E^{-2}$ spectrum as function of declination for different energy ranges.}
\label{fig:sensitivity_energy}
\end{figure}

Searches for point-like sources have the advantage that for a particular source they observe only a small portion of the sky (typical angular resolutions for muon neutrinos are around $0.5^\circ$) thereby reducing the background of atmospheric neutrinos considerably. Up to now, IceCube has observed about 180,000 up-going muon neutrinos in 1373 days of livetime \cite{arxiv:1406.6757}, ANTARES in a similar time about 5,500 \cite{apj:760:53}. None of the two experiments has found a significant deviation from the background, neither in searches on the whole sky nor for selected sources. 

This leads to upper limits on the muon neutrino flux from point-like sources which are plotted in Fig.~\ref{fig:ps_limits} as function of declination for an assumed $E^{-2}$ source spectrum. The comparison to the upper limits from MACRO \cite{apj:546:1038} illustrates the huge improvement in sensitivity of a factor 1000 over the past 15 years. Currently, IceCube is the most sensitive operating neutrino telescope. In the northern sky, its sensitivity starts to approach the \emph{discovery region} below $\sim10^{-12}\,\mathrm{TeV}\,\mathrm{cm}^{-2}\,\mathrm{s^{-1}}$ where according to calculations \cite{apj:656:870,app:34:778} fluxes from (Galactic) sources are expected. For sources in the southern sky, ANTARES located in the Mediterranean Sea is currently the most sensitive detector \cite{apj:760:53,apj:786:l5}, in particular in the TeV to PeV range, where the sensitivity of IceCube rapidly deteriorates due to the huge background of atmospheric muons (Fig.~\ref{fig:sensitivity_energy}). This energy region is in particular interesting for Galactic sources which are expected to emit neutrinos in this energy range. However, with its $\sim100$ times smaller volume, the overall sensitivity of ANTARES is much lower than that of IceCube. 

\subsection{Neutrinos from gamma-ray bursts}

\begin{figure}[t]
\includegraphics[width=8cm]{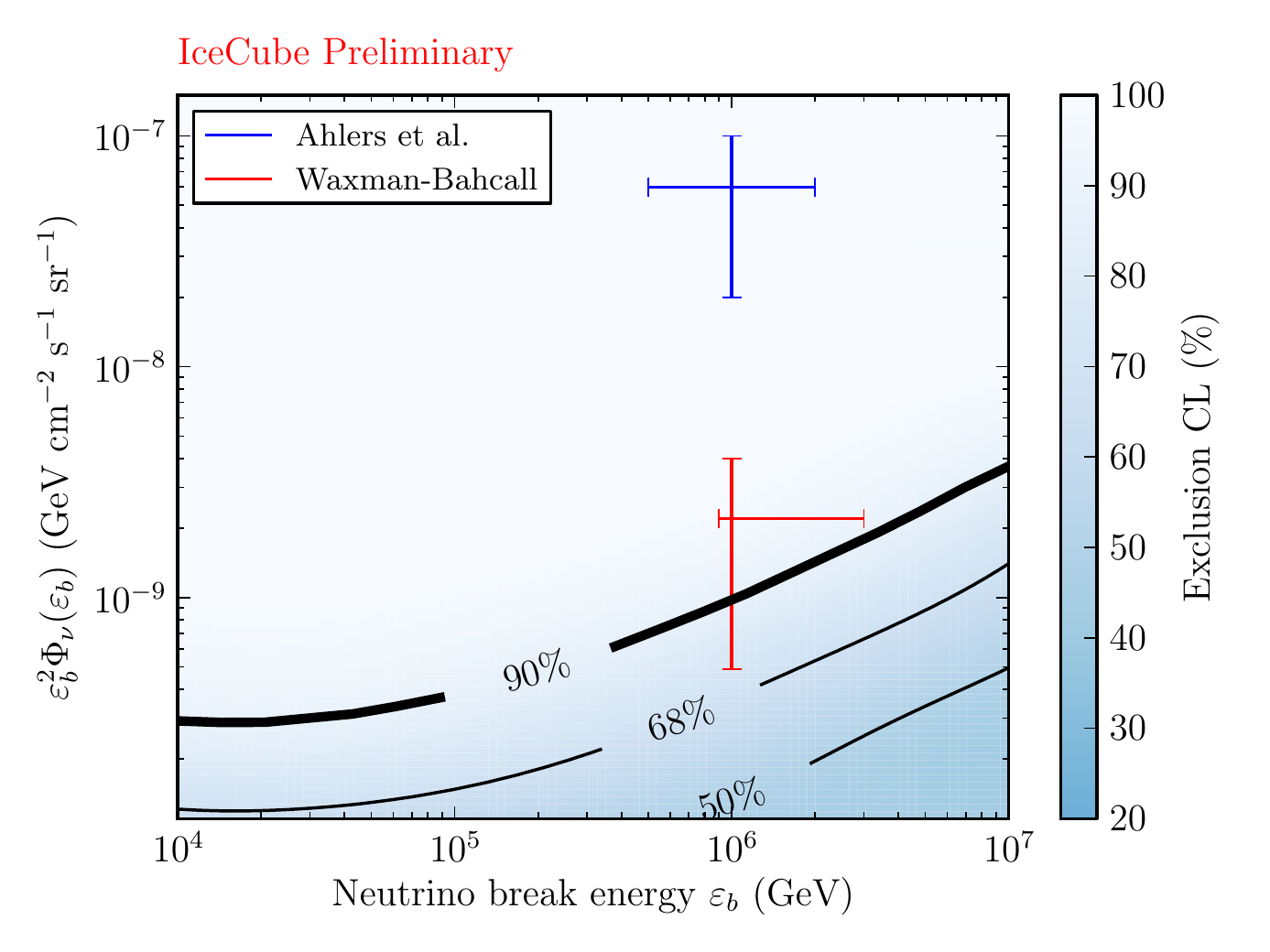}
\caption{Compatibility of neutrino flux predictions based on cosmic ray production in GRBs with observations. The horizontal axis shows the energy of the first break in the standard double-broken power law neutrino spectrum of GRBs in the fireball model, which is proportional to the square of the Lorentz factor of the ejected material. The vertical axis displays the normalization of the neutrino flux which is related to the accelerated proton flux. The area below the lines are the allowed values in the neutrino flux versus neutrino break energy plane at the corresponding confidence level. Model predictions \cite{prl:78:2292,app:35:87} with estimated uncertainties are shown as points.}
\label{fig:icecube_grbs}
\end{figure}

Apart from Active Galactic Nuclei (AGNs), Gamma-Ray Bursts (GRBs) are currently the only good candidates for the sources of the ultra-high energy cosmic rays above $10^{19}$\,eV. The fact that GRBs are transient phenomena with durations from below a second to several hundred seconds, allows for another significant reduction of the atmospheric background. 

The currently leading model for GRBs is the fireball model \citep{apj:405:278} with the energy source being the rapid accretion of a large mass onto a black hole formed after the collapse of a super-massive star. In this model, a highly relativistic outflow (fireball) dissipates its energy via synchrotron or inverse Compton radiation of electrons accelerated in internal shock fronts. This radiation in the keV--MeV range is observed as a broken power-law spectrum in gamma rays. In addition to electrons, protons are thought to be accelerated which interact with the keV--MeV photons via $p \gamma \rightarrow \pi^+, \pi^0$. The decaying charged pions produce neutrinos of energy ${\cal O}(10^{14}\,\mathrm{eV})$. Because of the $\Delta^+$ resonance, the neutrino spectrum mirrors the broken power-law shape of the gamma-ray spectrum with the break lying in the 100\,TeV range. The cooling of the pions at higher energies in magnetic fields introduces a second break in the PeV range.

Despite being the leading model, several of its parameters are still poorly determined or completely unknown, leading to significant uncertainties in the predicted fluxes. This situation is improved with so-called neutron escape models of GRBs \cite{prl:78:2292,app:35:87}. In these models, the neutrino flux is closely tied to the observed extragalactic cosmic-ray spectrum. Because of this direct link, these models are more robust against assumptions on unmeasured parameters. IceCube has searched for correlations of neutrinos with over 500 bursts, without success so far \cite{na:484:351} providing significant constraints on the models. Figure~\ref{fig:icecube_grbs} shows the exclusion limit and compares it to predictions from two neutron-escape models. Though not all models have been fully excluded yet within their uncertainties, the persistent null results significantly question the role of GRBs as major sources of the UHECRs. 

\begin{figure*}[t]
\includegraphics[width=7cm]{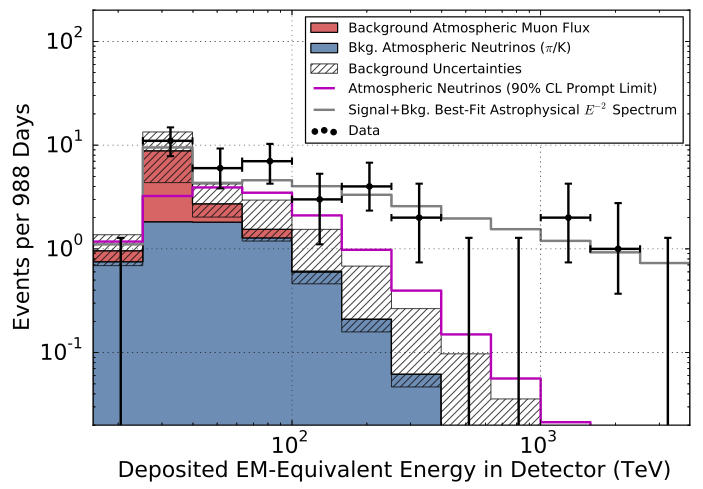}
\hfill
\includegraphics[width=9cm]{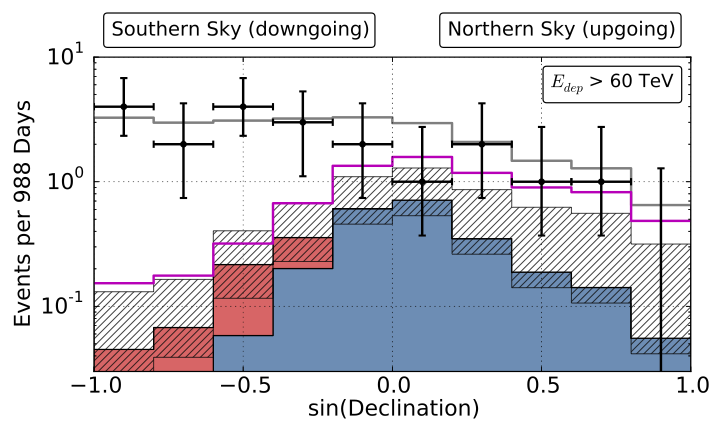}
\caption{High-energy neutrinos as observed by IceCube in 3 years of data. Left: Number of measured events and predicted fluxes as function of deposited energy equivalent to an electromagnetic shower. Right: Distribution of data and predictions as a function of declination for deposited energies $> 60$\,TeV (same color code as left). Taken from \cite{prl:113:101101}.}
\label{fig:hese}
\end{figure*}

\section{Neutrinos from above -- the power of veto}
For a long time, the field of view of neutrino telescopes was thought to be restricted to the opposite hemisphere for the observation of cosmic neutrinos in the TeV range. However, neutrinos interacting inside the instrumented volume can be separated from atmospheric muons, which enter the detector from the outside, by requiring that the outer layers of the detector do not contain any correlated signals. As atmospheric neutrinos are always produced together with muons in the atmosphere, this technique can even be used to suppress the flux of atmospheric neutrinos as first suggested in \cite{pr:d79:043009}. Depending on the size of the veto region, this technique allows for lowering the energy threshold for observations of down-going cosmic neutrinos significantly at the expense of restricting the effective volume for neutrino interactions to the non-veto volume of the detector. The latter could be avoided by using a detector at the surface like IceTop to reject atmospheric showers. With its current extension, however, the solid angle coverage of IceTop is rather limited and up to now it has only been used to check for the presence of an air shower in coincidence with an IceCube event in hindsight.

\section{Discovery of cosmic neutrinos with IceCube}

\begin{figure}[t]
\includegraphics[width=7.5cm]{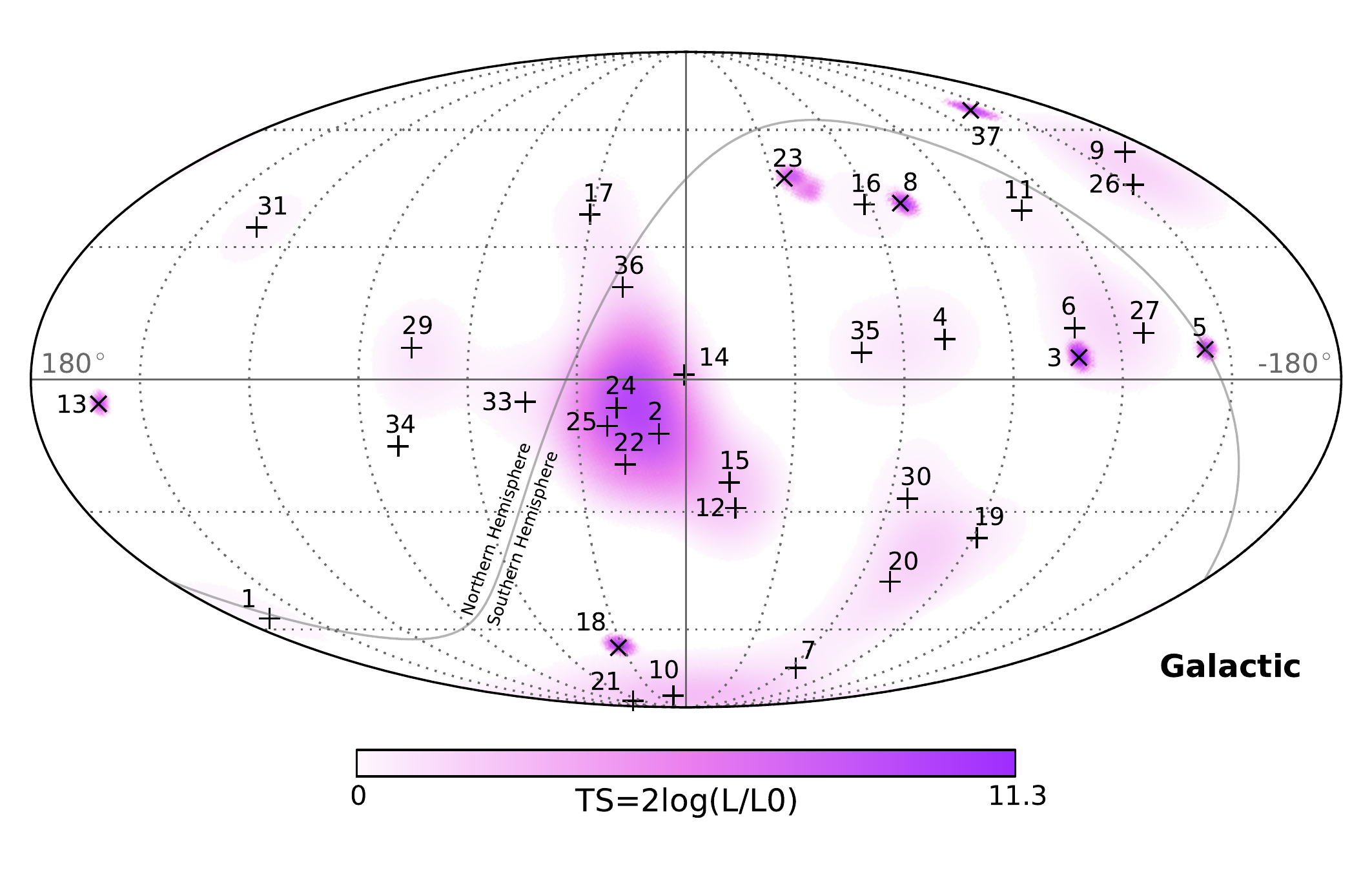}
\caption{Arrival directions of the events in Galactic coordinates of high-energy neutrinos observed by IceCube. Shower-like events are marked with $+$ (resolution $\sim 15^\circ$) and those containing muon tracks with $\times$ (resolution $\lesssim 1^\circ$). The Color scale indicates the compatibility with the background-only hypothesis for a point-like source. No significant clustering was observed with the most significant fluctuation having a p-value of $84\%$. Taken from \cite{prl:113:101101}.}
\label{fig:hese_skymap}
\end{figure}

\begin{figure}[t]
\includegraphics[width=7cm]{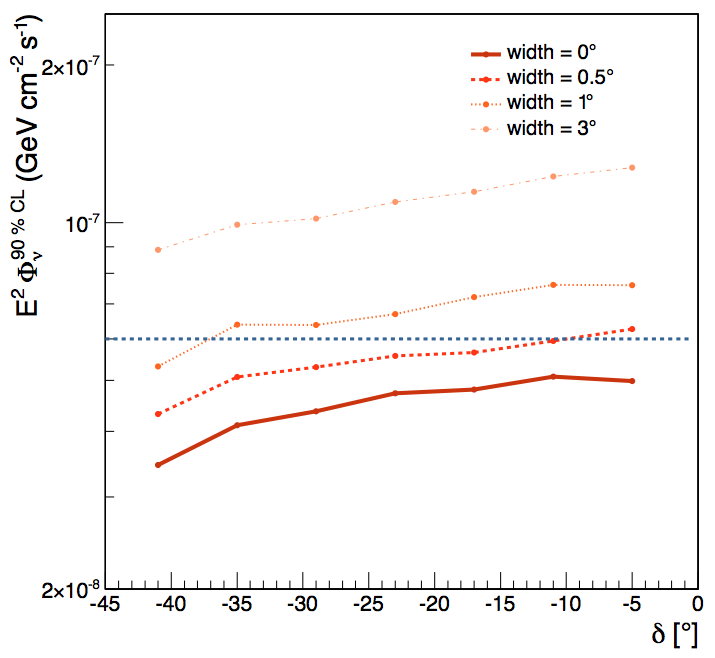}
\caption{ANTARES exclusion limit at 90\% CL.\ for a neutrino source near the Galactic Center assuming widths between $0^\circ$ (point source) and $3^\circ$. The blue horizontal line represents the source flux predicted in \cite{app:57:39}. Taken from \cite{apj:786:l5}.}
\label{fig:antares_gc_exclusion}
\end{figure}
\begin{figure}[t]
\includegraphics[width=8cm]{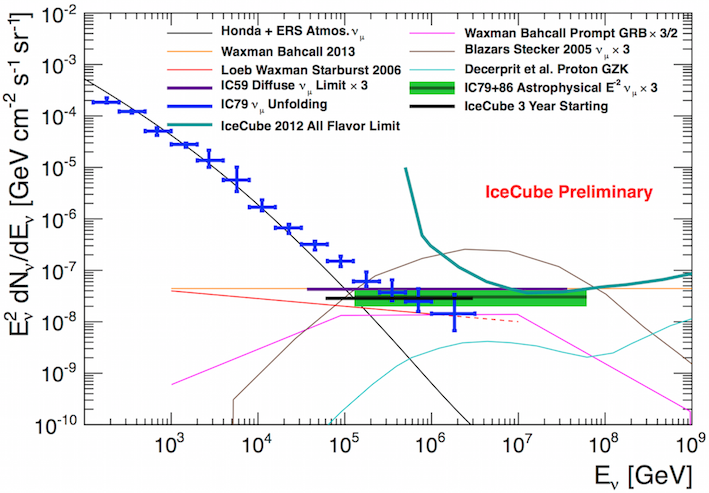}
\caption{Atmospheric and astrophysical neutrino fluxes measured by IceCube together with corresponding predictions. Taken from \cite{hill:neutrino:2014}.}
\label{fig:nu_fluxes}
\end{figure}

The veto method described above was developed and applied to search for cosmic neutrinos of all flavors in the IceCube detector. In two years of data, 28 events were observed over an expected background of $10.6^{+5.0}_{-3.6}$ yielding a significance of $4.2\,\sigma$ \cite{sc:342:1242856}. Adding another year of data provided 9 additional events and raised the rejection of the background-only hypothesis to $5.7\,\sigma$ \cite{prl:113:101101}. Three of the altogether 37  events deposited more than 1\,PeV of energy inside the detector \footnote{Note, that this energy corresponds to the amount of light deposited by an electromagnetic cascade of that energy, and in particular does not account for the energy carried away by the neutrino in a NC interaction or by the muon when it leaves the detector.}. Figure~\ref{fig:hese} (left) shows the distribution of observed and expected events as a function of deposited energy. A clear excess at high energies can be observed which can be accounted for with an astrophysical component. In the northern sky, the flux of cosmic neutrinos is suppressed due to Earth absorption whereas in the southern sky it stays constant and is clearly separated from the atmospheric neutrino flux which is suppressed by the veto (Fig.~\ref{fig:hese}, right). 

A combined fit of a conventional, prompt and $E^{-2}$ astrophysical neutrino flux to the data between 60\,TeV and 3\,PeV yields a best-fit astrophysical normalization of $E^2 \phi = (0.95 \pm  0.3) \times 10^{-8}\,\mathrm{GeV}\,\mathrm{cm}^{-2}\,\mathrm{sr}^{-1}$; the best-fit slope is $E^{-2.3\pm0.3}$ with a normalization of $E^2 \phi = 1.5 \times 10^{-8}\,(E/100\,\mathrm{TeV})^{-0.3}\,\mathrm{GeV}\,\mathrm{cm}^{-2}\,\mathrm{sr}^{-1}$ \cite{prl:113:101101}. The observed events are compatible with an isotropic neutrino flux with a flavor ratio ($\nu_e$:$\nu_\mu$:$\nu_\tau$) of (1:1:1) as expected from full mixing due to neutrino oscillations. This suggests that the observed cosmic neutrino flux is to a large extent of extra-galactic origin with possibly a component from the Milky Way's halo. 

Searches for a directional (Fig.~\ref{fig:hese_skymap}) or temporal clustering of the events as well as correlations with the Galactic plane yield negative results \cite{prl:113:101101}. Though statistically not significant, a cluster of seven events near the Galactic Center inspired flux calculations for a potential associated source \cite{app:57:39}. The ANTARES Collaboration tested this hypothesis and excludes the existence of such a source with an extension of up to $0.5^\circ$ with more than 90\% CL \cite{apj:786:l5} (Fig.~\ref{fig:antares_gc_exclusion}). 

In the meantime, cosmic neutrinos have also been seen with the IceCube detector in up-going muon neutrinos: in two years of data an analysis that searched for a diffuse cosmic neutrino flux observed an excess of $3.9\,\sigma$ (preliminary) at high energies over the predicted atmospheric background. The best-fit astrophysical component (green horizontal band in Fig.~\ref{fig:nu_fluxes}) is compatible in slope and normalization with the cosmic neutrino flux discussed above. 

\section{Future strategies}
Currently, only the IceCube detector provides the required volume to investigate the observed cosmic neutrino flux. Though the full potential of reconstruction and selection methods has not been fully exploited yet, it is foreseeable that even after 10 years of operation, IceCube will only have gathered about 90 astrophysical muon neutrinos above 100\,TeV, 100 cascades above 60\,TeV and 10 events above 1\,PeV. For a detailed investigation of the properties of the astrophysical flux and in particular the search for individual sources of neutrino emission, significantly more events will be required. Accordingly, the collaborations of all three running neutrino telescopes are either working on successor experiments or on upgrading their detectors to the gigaton scale. 

In order to coordinate these efforts, an umbrella organization, the Global Neutrino Network (GNN), was founded in 2013 \cite{url:gnn}. It includes all current neutrino telescope projects ANTARES, BAIKAL, IceCube and KM3NeT and aims at developing a coherent strategy for the field. This covers o.a.\ to understand how the different detectors complement each other in the best way, the coordination of alert and multi-messenger policies, development of standards for data exchange, cross-checks of results as well as the organization of meetings for exchange of expertise.

\subsection{Next generation IceCube}

\begin{figure}[t]
\includegraphics[width=7cm]{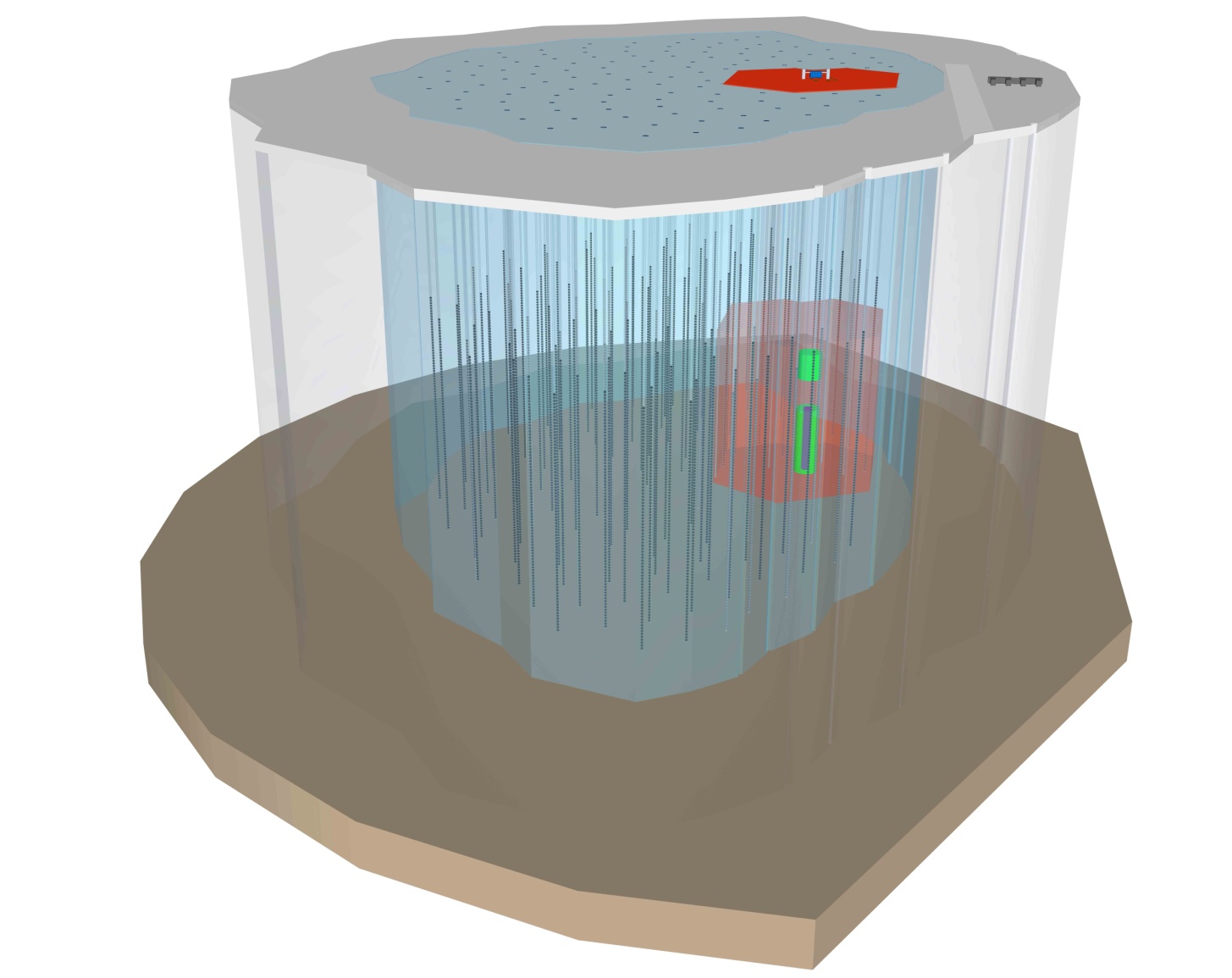}
\caption{Artist's view of a high-energy (blue volume) and low-energy (green volume) extension of the IceCube neutrino detector. The current IceCube detector is shown in red.}
\label{fig:icecube_ng}
\end{figure}

\begin{figure}[t]
\includegraphics[width=8cm]{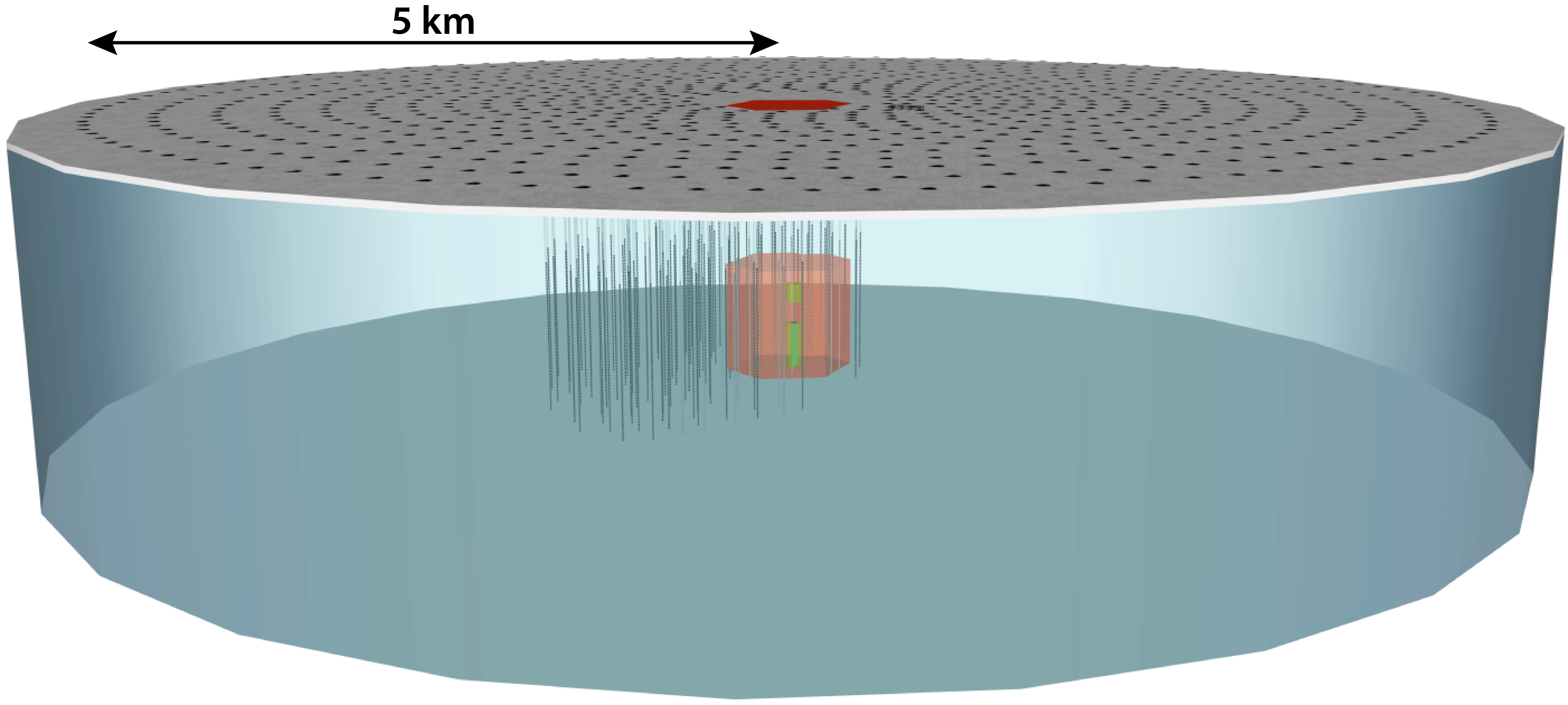}
\caption{Artist's view of a surface veto for the IceCube neutrino detector.}
\label{fig:iceveto_1}
\end{figure}

\begin{figure}[t]
\center
\includegraphics[width=7cm]{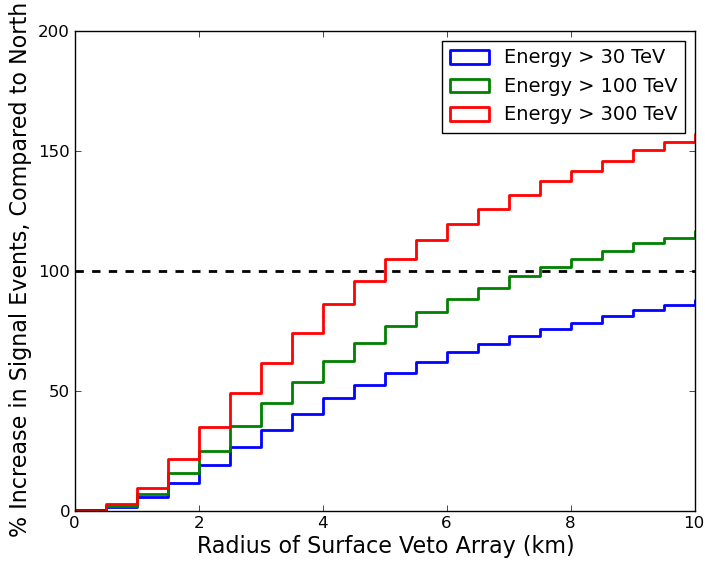}
\caption{Increase in muon neutrino events for the current IceCube detector compared to purely up-going muon neutrinos as a function of the radius of a 100\% efficient surface veto for various energy thresholds. For the neutrino spectrum the best-fit cosmic neutrino flux with an index of $-2.3$ and no high-energy cutoff was assumed. Taken from \cite{karle:nbic:2014}.}
\label{fig:iceveto_2}
\end{figure}

The IceCube Collaboration is currently planning an extension of the existing detector with about 120 additional strings, each equipped with up to 96 optical modules. An artist conception of such an extended IceCube detector is depicted in Fig.~\ref{fig:icecube_ng}. Intensive studies for optimizing this future detector are underway. 80 of these strings are foreseen for a high-energy extension with a wide horizontal spacing of up to 300\,m, yielding instrumented volumes between 5 and 10\,km$^3$ and angular resolutions down to $0.1^\circ$ with an energy threshold in the some 10\,TeV range.  

The remaining 40 strings are planned to be installed in a very dense configuration called PINGU in the center of the currently existing detector to allow for the reconstruction of neutrinos in the GeV range. The main goal of PINGU is the measurement of the neutrino mass hierarchy exploiting matter effects in the oscillation pattern of atmospheric neutrinos (for details see \cite{arxiv:1401:2046}). 

\begin{figure*}[t]
\includegraphics[width=7cm]{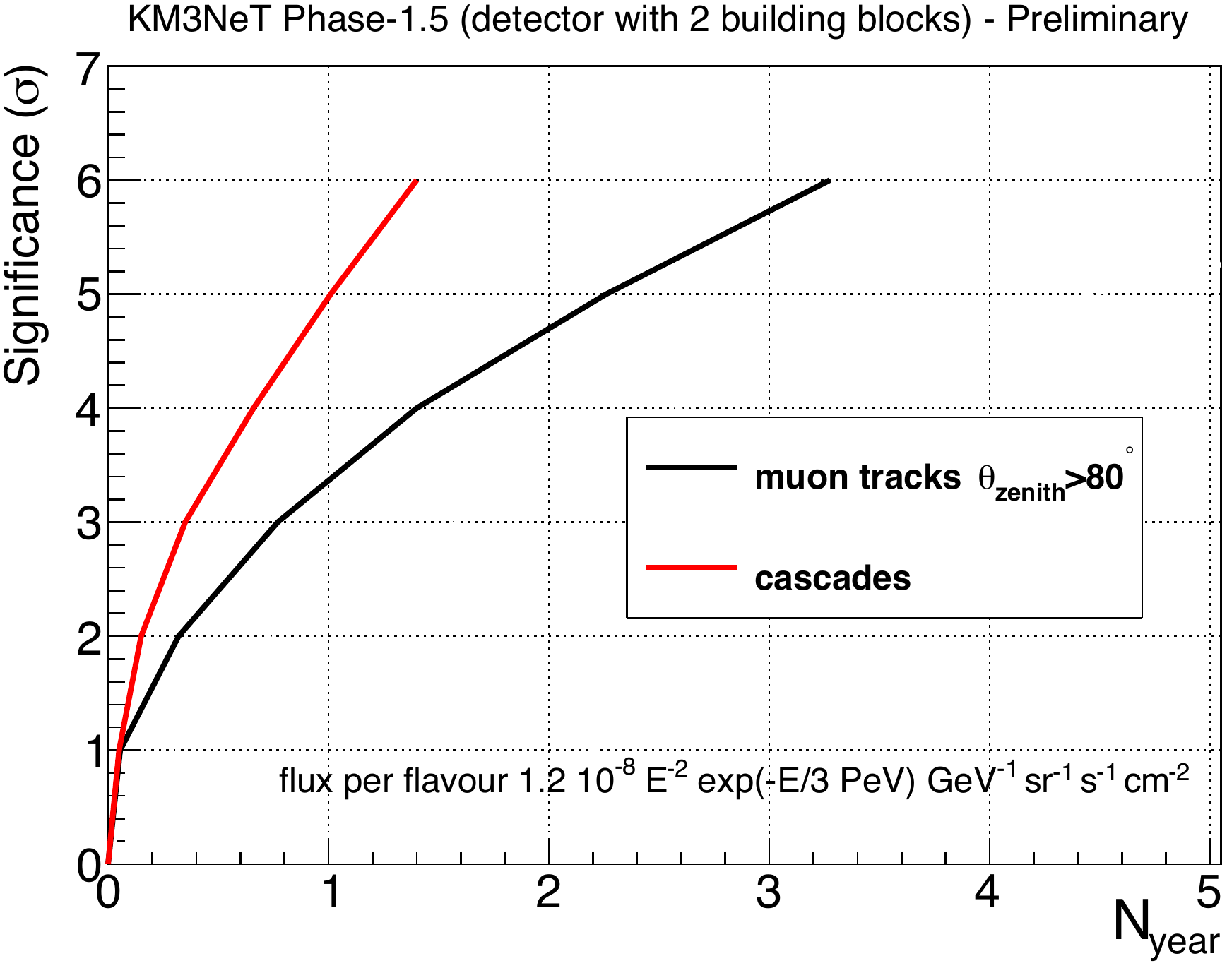}
\hfill
\includegraphics[width=8cm]{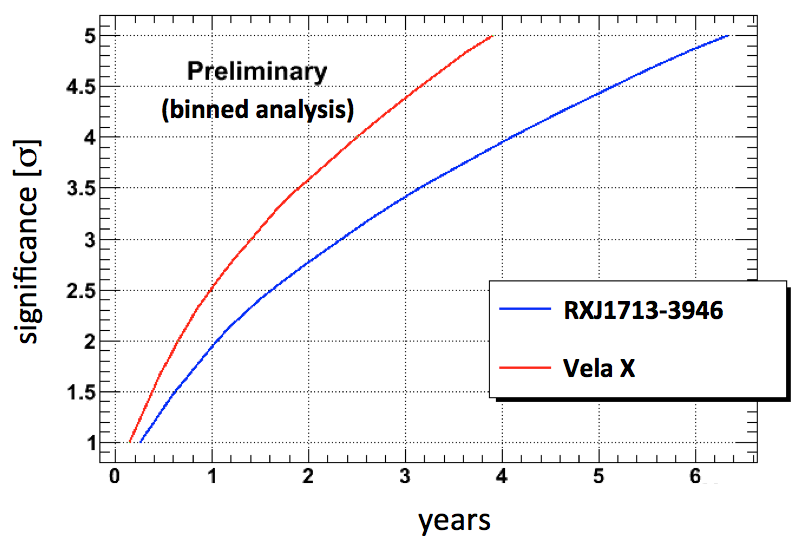}
\caption{Left: Sensitivity of KM3NeT with two building block (115 lines each) to the cosmic neutrino flux discovered by IceCube as a function of observation time for muon tracks and cascades. Taken from \cite{james:tmex:2014}. Right: Sensitivity of KM3NeT with six building blocks to the supernova remnant RX\,J1713.7-3946 and the pulsar wind nebula VelaX as a function of observation time assuming that the observed gamma-ray flux fully originates from $\pi^0$ decay. Taken from \cite{dejong:neutrino:2014}.}
\label{fig:km3net}
\end{figure*}

In addition to the in-ice extensions, the potential of an extended IceTop-like detector for vetoing atmospheric muons and neutrinos at the surface via the registration of the associated air shower is being investigated (see Fig.~\ref{fig:iceveto_1}). Such a surface veto has the virtue that the whole ice volume between the deep detector and the surface would be available for neutrino interactions thereby significantly enlarging the effective volume. In order to cover a reasonably large solid angle, such a detector would have to extend horizontally significantly further than the detector in the deep ice. The energy threshold is determined by the density of detection elements which can be very simple compared to the current IceTop detectors as only basic signal information is required. A first preliminary study concerning the gain in signal muon neutrinos with such a surface veto for the current IceCube detector is depicted in Fig.~\ref{fig:iceveto_2}. For a realistic energy threshold of 100\,TeV and an extension radius of 5\,km, an increase of about 75\% compared to up-going muon neutrinos can be expected.

\subsection{KM3NeT}
The KM3NeT Collaboration, which comprises the expertise of all previous Mediterranean neutrino telescope projects (ANTARES, NEMO, NESTOR), intends to build a multi-km$^3$ neutrino telescope in the Mediterranean Sea \cite{km3net:tdr:2010}. For the optical module, it uses an innovative design \cite{nim:a718:513} with 31 small 3" photomultipliers instead of a single large (8--10") photomultiplier as adopted in current neutrino telescopes. Among others, this configuration yields three times the sensitive area of an optical module with a single 10" photomultiplier, provides intrinsic directional sensitivity and allows for improved photon counting. The full detector will contain over 12,000 of these sensors distributed over about 600 vertical lines arranged in 6 so-called building blocks of 115 lines each. It is planned to install these building blocks at sites near France, Italy and Greece. A single building block will allow to reconstruct the direction of muon neutrinos with resolutions down to $0.1^\circ$ at high energies. Furthermore, studies show that a precision of $3^\circ$ in the direction reconstruction of cascades can be reached at 10\,TeV, and better than $2^\circ$ above 100\,TeV. This would for the first time allow to also use the cascade channel in searches for point-like neutrino sources, thereby significantly increasing the number of events for a given flux.

A first phase of KM3NeT with 31 strings has been funded with 31\,M\euro\ and will be installed until the end of 2016 together with 8 towers containing ''traditional'' optical modules with one large PMT per module. The main purpose of this first phase is the demonstration of the functionality of the detector design but it will also provide improved performance over the ANTARES detector. Afterwards, the construction of two building blocks is envisaged (additional costs 50--60\,M\euro). Such a configuration would have about the same instrumented volume as the current IceCube detector with the main purpose to verify and further investigate the observed cosmic neutrino flux. Figure~\ref{fig:km3net} (left) displays a preliminary estimate of the sensitivity in tracks and cascades to this flux. With the full detector in the final phase (additional costs 140--160\,M\euro), KM3NeT will instrument between 3 and 6\,km$^3$ of water and be sensitive to Galactic neutrino sources like the supernova remnant RX\,J1713.7-3946 or the pulsar wind nebula Vela\,X (Fig.~\ref{fig:km3net}, right). The KM3NeT Collaboration currently also investigates the performance of a densely instrumented detector called ORCA consisting of one building block. As with PINGU, the main goal is the measurement of the neutrino mass hierarchy.

\subsection{Baikal}
The Baikal Collaboration plans the stepwise installation of a km$^3$-scale array in Lake Baikal dubbed GVD (Gigaton Volume Detector) \cite{nim:a742:82}. It will consist of clusters of 8 strings with up to 48 optical modules per string. Phase 1 of the project envisages 12 clusters. In 2008–-2013, the basic elements were tested and engineering strings operated. Meanwhile, the first cluster is deployed to a large part and completion is planned for April 2015.

\section{Summary}
The discovery of an astrophysical neutrino flux with the IceCube detector has provided a strong boost for neutrino astronomy and the community is currently working intensively on fully exploiting the available and upcoming data as well as understanding the nature and sources of these neutrinos. However, it is quite clear that even after 10 years of running with the current detectors, only a limited number of astrophysical neutrinos will have been collected, likely not sufficient to answer the many questions that have emerged. In particular, it will probably be difficult to identify the sources of the Galactic and extra-galactic cosmic rays.

Therefore, in order to exploit the full potential of neutrino astronomy, a next generation of neutrino telescopes with instrumented volumes in the 5--10\,km$^3$ range with good vetoing efficiency for atmospheric muons and neutrinos and full sky coverage in up- and down-going neutrinos is required. The planning and performance studies for these detectors are currently in full swing and, pending funding, could be realized within 10--15 years. With these new detectors we will then hopefully be able to fully open this new and exciting window to the universe.

\section{Acknowledgements}
The author gratefully acknowledges the support by the German Ministry for Education and Research (BMBF), grant numbers 05A11KHB and 05A14WE3.











\end{document}